\documentclass[fleqn,usenatbib]{mnras}

\usepackage{newtxtext,newtxmath}

\usepackage[T1]{fontenc}
\usepackage{lineno}
\usepackage{hyperref}
\usepackage{graphicx}	
\usepackage{dblfloatfix}
\usepackage{amsmath}	
\usepackage{threeparttable}
\usepackage{subcaption}
\usepackage{caption}
\usepackage{threeparttable}
\usepackage{color}
\usepackage[dvipsnames]{xcolor}
\usepackage[normalem]{ulem}
\usepackage{booktabs}
\usepackage{multirow}
\def \link_col{blue}
\DeclareRobustCommand{\VAN}[3]{#2}
\let\VANthebibliography\thebibliography
\def\thebibliography{\DeclareRobustCommand{\VAN}[3]{##3}\VANthebibliography}

\usepackage{graphicx}	
 
\usepackage{amsmath}	
\usepackage{floatrow}
\usepackage{mathtools} 

\usepackage{graphicx}	
\usepackage{amsmath}
\usepackage{amssymb}
\usepackage{amsfonts}
\usepackage[utf8]{inputenc}

\title[]{Detection of two TeV gamma-ray outbursts from NGC 1275 by LHAASO}

\author[LHAASO Collaboration]{\parbox{\textwidth}{
\normalsize
Zhen Cao$^{1,2,3}$,
F. Aharonian$^{4,5}$,
Axikegu$^{6}$,
Y.X. Bai$^{1,3}$,
Y.W. Bao$^{7}$,
D. Bastieri$^{8}$,
X.J. Bi$^{1,2,3}$,
Y.J. Bi$^{1,3}$,
W. Bian$^{9}$,
A.V. Bukevich$^{10}$
% ...
(LHAASO Collaboration)
}
}

\date{Accepted XXX. Received YYY; in original form ZZZ}

\pubyear{2024}

\DeclareUnicodeCharacter{FF0C}{,}
\begin{document}
\maketitle
\label{firstpage}
\pagerange{\pageref{firstpage}--\pageref{lastpage}}

\begin{abstract}
The Water Cherenkov Detector Array (WCDA) is one of the components of Large High Altitude Air Shower Observatory (LHAASO) and can monitor any sources over two-thirds of the sky for up to 7 hours per day with >98\% duty cycle. 
In this work, we report the detection of two outbursts of the Fanaroff-Riley I radio galaxy NGC 1275 that were detected by LHAASO-WCDA between November 2022 and January 2023 with statistical significance of 5.2~$\sigma$ and 8.3~$\sigma$. The observed spectral energy distribution in the range from 500 GeV to 3 TeV is fitted by a power-law with a best-fit spectral index of $\alpha=-3.37\pm0.52$ and $-3.35\pm0.29$, respectively.
The outburst flux above 0.5~TeV was 
($4.55\pm 4.21)\times~10^{-11}~\rm cm^{-2}~s^{-1}$ and 
($3.45\pm 1.78)\times~10^{-11}~\rm cm^{-2}~s^{-1}$, corresponding to 60\%, 45\% of Crab Nebula flux.
Variation analysis reveals the variability time-scale of days at the TeV energy band. 
A simple test by one-zone synchrotron self-Compton model reproduces the data in the gamma-ray band well. 
\end{abstract}



\section{INTRODUCTION}


Relativistic jets launched from active galactic nuclei (AGNs) represent one of the most intriguing classes of very-high-energy (VHE) extragalactic sources.
Investigating VHE emission from relativistic jets offers insights into the acceleration of cosmic rays in large-scale structures and the interactions of relativistic particles with the intergalactic medium. NGC 1275, also known as Perseus A, is a prominent galaxy located in the galaxy cluster Perseus at a redshift 0.0176 \citep{1995ApJS...98..219Y}. It is classified as a radio-loud Seyfert 1.5 AGN \citep{1997ASPC..113..429H}. Notably, NGC 1275 exhibits intriguing large-scale morphology, particularly in its jet interaction with the surrounding medium at the pc length scale. \citep{2021ApJ...920L..24K,2022MNRAS.509.1024O}. 

NGC 1275 has been observed across various wavelengths, from radio to gamma-ray energy bands, and a long-term monitoring has been carried out \citep{2018ApJ...860...74T,2021ApJ...914...43H,2022ApJ...925..207Z,2023A&A...669A..32P}. 
In the gamma-ray band, NGC 1275 was detected by the Fermi Large Area Telescope (Fermi-LAT). During the first four months of all-sky-survey observations \citep{Fermi-LAT:2009mxo}, NGC 1275 was found to have an average flux of $F = (2.10\pm 0.23) \times 10^{-7}~\rm ph\, cm^{-2} s^{-1}$ above $100~\rm MeV$ and a power-law index of $(-2.17\pm 0.05)$. The GeV emission is variable on a scale of a few months. 
\citep{2010ApJ...715..554K,2011MNRAS.413.2785B,2018ApJ...860...74T}. Several flares have been detected \citep{2010ATel.2737....1D}, indicating that only a subdominant fraction of the detected GeV emission comes from the galaxy cluster. 
An analysis of 8 years of Fermi-LAT gamma-ray data for NGC 1275 revealed a gradual increase in the flux level together with significant short-term flux variability. The spectral behavior changed notably around February 2011. Initial changes were likely due to high-energy electron injections into the jet, while later flares may have resulted from variations in the jet Lorentz factor or the jet direction \citep{2018ApJ...860...74T}.

The VHE emission of NGC 1275 was detected by the Major Atmospheric Gamma Imaging Cherenkov (MAGIC) telescope system with a statistical significance of $6.1\sigma$ between October 2009 and February 2010, and $6.6\sigma$ from August 2010 to February 2011 \citep{2012A&A...539L...2A}. The measured differential energy spectrum can be described by a power-law with a large spectral index of $(-4.1\pm0.7_{\rm stat}\pm0.3_{\rm syst})$. 
Observations by Fermi-LAT and MAGIC suggest that the ~\rm GeV--TeV emission region of NGC 1275 is not near the central massive black hole \citep{2021MNRAS.500.4671L}, indicating that the VHE emission might originate from the relativistic jet or the region of its interaction with the surrounding medium. 

Recently, NGC 1275 exhibited flaring activity on December 20-22, 2022, as reported by the MAGIC collaboration (2022, ATel\#15820) \citep{2022ATel15820....1B}. The statistical significance of detection exceeded a remarkable value of 30 standard deviations, and the flux level was comparable to that of the Crab Nebula. This flux level surpassed that detected by MAGIC in 2010 by a factor of several tens and rivaled the flux observed by MAGIC during the notable flaring state from December 2016 to January 2017. Additionally, the Major Atmospheric Cherenkov Experiment (MACE) has detected another flaring episode from NGC 1275 on 2023 January 10-11. Preliminary data analysis using MAP (MACE data Analysis Package) indicated an outburst with a statistical significance of $\sim$7.87$\sigma$, with the flux reaching approximately 1 Crab unit (2023, Atel\#15856) \citep{2023ATel15856....1Y}.

The sensitivity and wide-field-view capabilities of Water Cherenkov Detector Array (WCDA), one sub-array of The Large High Altitude Air Shower Observatory (LHAASO) have successfully enabled the detection of low-luminosity AGN NGC 4278 \citep{2024arXiv240507691C}. In this paper, based on unprecedented long-term TeV light curve we will focus on two clear outburst episodes of NGC 1275 that LHAASO-WCDA detected between November 2022 and January 2023. 

The paper is structured as follows. In Section 2, we provide a brief introduction to the LHAASO experiment and discuss the production of sky maps. Section 3 covers the analysis methods and related results, including the application of forward-folding analysis to estimate flux and spectral measurements, as well as the algorithms utilized for characterizing variability. Additionally, we present the multi-wavelength results obtained from ZTF, MAXI-GSC and Fermi-LAT during the same observation period. A one-zone SSC model is used to fit the VHE dath. Finally, Section 4 offers a summary.

\section{THE LHAASO Data, Event, and Background Maps}

LHAASO, situated at Mt. Haizi at an altitude of 4410 meters above sea level in Daocheng, Sichuan Province, China, is a multifaceted and comprehensive extensive air shower detector array. LHAASO has been strategically designed to facilitate research on cosmic rays and gamma rays across a broad energy spectrum ranging from hundreds of GeV to over a PeV. The facility comprises three main sub-arrays: the $1.3~\rm {km}^{2}$ ground array (KM2A), the 
WCDA, and the Wide Field-of-view Air Cherenkov/Fluorescence Telescope Array (WFCTA). Detailed information on the different detectors, their performance, and data reconstruction algorithm is extensively discussed in 
\citet{2024ApJS..271...25C}.

The dataset utilized in this analysis was mainly sourced from LHAASO-WCDA. The duration for data collection spans from November 28, 2022, to January 29, 2023. For this study, we restricted our analysis to events with a zenith angle of less than 50 degrees to ensure high-quality reconstruction data. 
Additionally, a gamma/proton separation parameter of less than 1.1 is employed to effectively identify gamma-like events.
The selected events are subsequently categorized into six groups based on the effective number of triggered PMT units, denoted as $N_{hit}$, with ranges [60-100), [100-200), [200-300), [300-500), [500-800), and [800-2000). 
The median energies for these groups are 0.59, 1.11, 2.16, 3.76, 6.84, and 12.20 TeV, respectively. Ultimately, the effective live-time for observations of NGC 1275 was recorded as 60 transits, with an approximate count of $3\times10^8$ gamma-like events. 

To calculate the excess signal from the source, six event and background maps relative to above six $N_{hit}$ are generated using WCDA data within a $10^{\circ}\times 10^{\circ}$ area with a grid of $0.1^{\circ}\times0.1^{\circ}$, centered at the NGC 1275 nominal position. The event maps are created as histograms of the arrival direction of the reconstructed events, after converting the arrival direction of the shower from local coordinates to equatorial coordinates. The background maps are computed using a direct integration method \citep{2004ApJ...603..355F}. For this analysis, a sliding time window of 4 hours was used to estimate the detector acceptance, with an integration time set to 1 hour to estimate the relative background. Events within the region of the Galactic plane ($|b| < 10^{\circ}$) and LHAASO 1st gamma-ray source catalog \citep{2024ApJS..271...25C} (with a spatial angle less than 5 degrees) were excluded from the background estimation. The excess map is then obtained by subtracting the background from the event data map.

\section{DATA ANALYSIS AND RESULTS} 

 
\subsection{Multi-band Light Curves} 

This section offers an overview of the available datasets, summarizing the operational energy range for each detector and outlining the data processing steps. Additional details can be found in the references provided in the corresponding subsections.

\begin{figure*}
\centering
\includegraphics[width=0.7\linewidth,height=12cm]{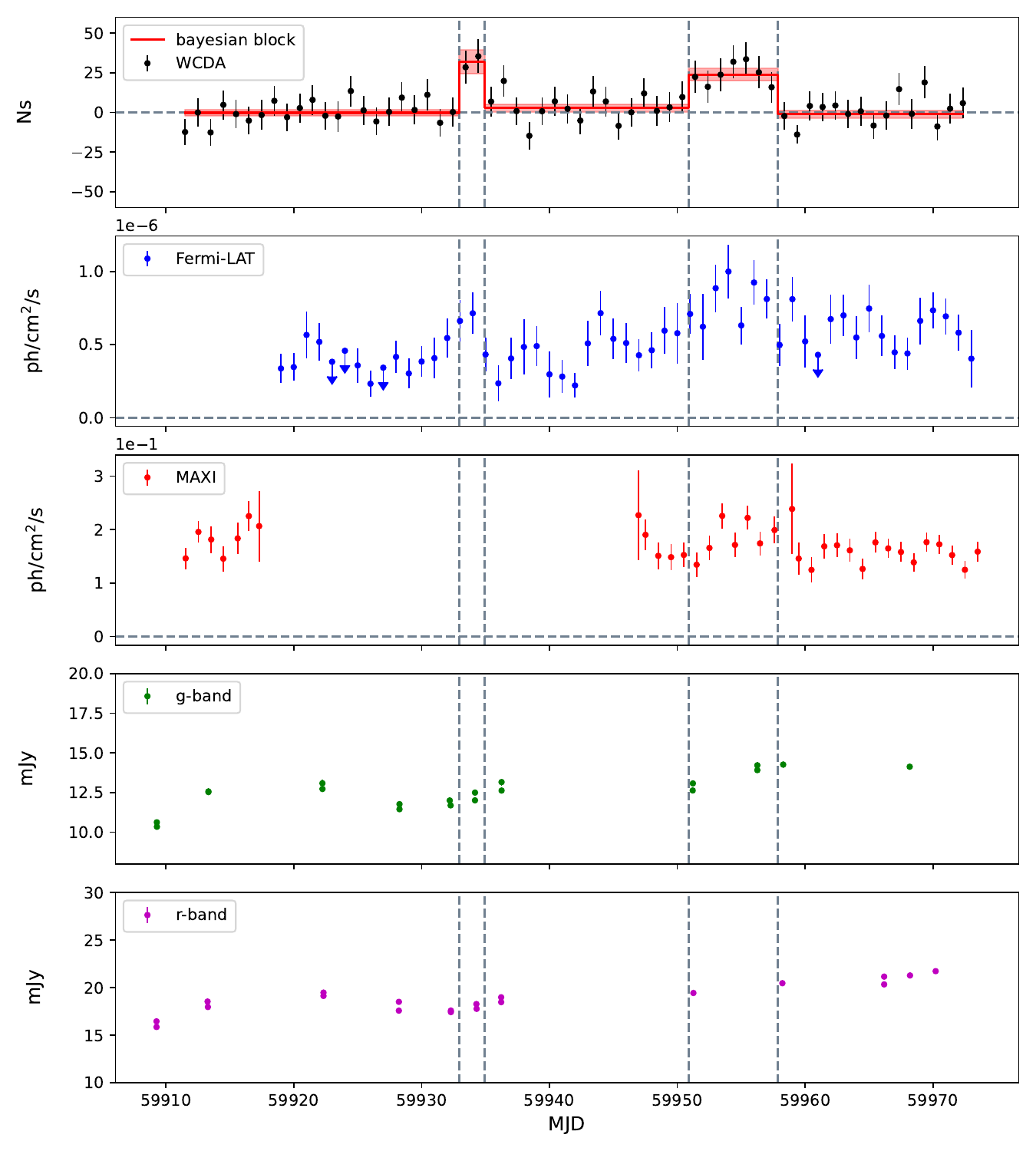}
\caption{Daily light curves of NGC 1275 in multiple energy bands from November 28, 2022, to January 29, 2023.  From top to bottom: LHAASO-WCDA data for $N_{hit} > 100$, Fermi-LAT data spanning 100~\rm MeV to 300~\rm GeV, MAXI-GSC data from 2~\rm keV to 20~\rm keV,  ZTF data in the g-band, and ZTF data in the r-band. Gray horizontal lines indicate zero flux levels, gray vertical lines mark two notable outburst periods. The red lines in top panel represent distinct $N_{s}$ states between change points identified via Bayesian blocks analysis, with a 10\% false positive probability.} 
\label{fig_mwl}
\end{figure*}

\subsubsection{VHE light curve}

The VHE light curve from LHAASO-WCDA is displayed in the top panel of Figure \ref{fig_mwl}. This light curve is obtained by using events with $N_{hit} > 100$, corresponding to a median gamma-ray energy of 1 TeV, assuming a Crab-like spectrum. The parameter $N_s$ represents daily signal events extracted from the previously described excess map. To accommodate variations in observation time and detection efficiency, a weighting factor is introduced to adjust $N_s$, ensuring the daily light curves accurately reflect the actual excess emission from the source. Specifically, the scaled $N_s = N_{bk, mean} \times ({N_{on}-N_{bk}})/{N_{bk}}$, where $N_{on}$ is the total number of events within the Point Spread Function (PSF) region, $N_{bk}$ is the background events in the same region for a single transit, and $N_{bk, mean}$ is the two-month average, set here as 80.52.

\subsubsection{HE light curve }

Fermi-LAT is designed to cover the energy band from $20~\rm MeV$ to greater than $300~\rm GeV$ \citep{2009ApJ...697.1071A}. 
The data surrounding NGC 1275 were obtained from Fermi Science Support, accounting for all sources within a 15-degree radius and standard cuts were applied to select the good time intervals (Z\_max \textless $90^\circ$, DATA\_QUAL \textgreater 0 and LAT\_config == 1). To investigate the gamma-ray flux variations of NGC 1275, the light curve variation in the energy band of $100~\rm MeV$ - $300~\rm GeV$ 
was calculated using a binned likelihood analysis. The appropriate instrument response function for this data set is P8R3\_SOURCE\_V3. The Galactic diffuse emission model (gll\_iem\_v07.fits) was used in the likelihood analysis, and the model for the extragalactic isotropic diffuse emission was iso\_P8R3\_SOURCE\_V3\_v1.txt. 
The spectral parameters are kept free for the sources within 5 degree, while are fixed to the values given in the 4FGL-DR2 \citep{2020ApJS..247...33A} catalog for other sources. The resulting integral flux in the energy range is shown in the middle panel of Figure \ref{fig_mwl}. All the time bins in which the Likelihood fit returned a TS <9 are shown as upper limits.

\subsubsection{ X-ray light curve}
 
The Monitor of All-sky X-ray Image (MAXI) onboard the International Space Station (ISS) has been monitoring the entire sky continuously since August 2009 \citep{2009PASJ...61..999M}. The light curves for specific sources are public \href{http://maxi.riken.jp/}{\textcolor{blue}{available}}. In this work, we utilized the MAXI on-demand service to compute the light curve for NGC 1275, using data from the MAXI-GSC, which operates in the 2-20 ~\rm keV energy band and covers 97\% of the sky daily. 

\subsubsection{Optical light curve}

The Zwicky Transient Facility (ZTF) is a new time-domain survey that had the first light at Palomar Observatory in 2017. Optical magnitudes in $g$ and $r$ bands were collected from the 20th ZTF public data release \citep{2019PASP..131a8003M}. The optical data during two outbursts were selected. After correction for Galactic extinctions (NED), the average magnitudes from each outburst were converted into data points on the SED.

\subsubsection{Bayesian Blocks}

In Figure \ref{fig_mwl}, two distinct outburst states are clearly observed 
in the ~\rm TeV and ~\rm GeV energy bands.
To quantitatively determine the start and end times of these light curve segments, the Bayesian blocks algorithm \citep{2013ApJ...764..167S} is applied to identify the optimal data segmentation. We utilize the point measurement fitness function as specified in the algorithm, configuring it with a false positive rate of 10\%.

This analysis revealed four change points and five blocks. 
Two distinct outbursts were identified: the first, O1, from Modified Julian Date (MJD) 59933.445 to 59935.439, and the second, O2, from MJD 59951.396 to 59958.376. 
These results are in agreement with the observations reported in ATel\#15820 and ATel\#15856, as illustrated by the vertical red lines in the top panel of Figure \ref{fig_mwl}.

Moreover, an apparent correlation between Fermi GeV data and WCDA TeV data is evident, as indicated by the positive correlation discussed in Section 3.3. 

\subsection{VHE variability analysis}

\begin{figure*}
\centering
\includegraphics[width=0.7\linewidth,height=5cm]{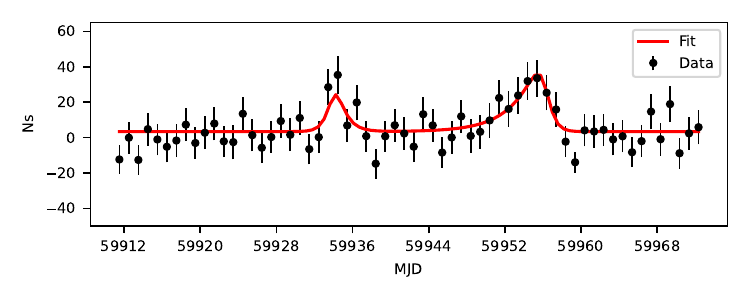}
\caption{LHAASO-WCDA light curve for $N_{hit} > 100$ with temporal fitting results.}
\label{fig_fred}
\end{figure*}

The so-called Fred function \citep{2010ApJ...722..520A} is employed to fit the WCDA light curve for a quantitative analysis of its time-varying behavior.
Figure \ref{fig_fred} displays the fitting results, in which the red line represents the fitting curve 
\begin{equation}
   F(t) = F_{c} + F_{0} \times \left[e^{(t_{0}-t)/t_{r}}+e^{(t-t_{0})/t_{d}}\right] ^{-1}\,, 
\end{equation}
where $F_{c}$ represents the constant level, and $F_{0}$ denotes the variability amplitude both measured in unit of $N_{s}$. The peak time in MJD is $t_{0}$, while the rise time $t_{r}$ and decay time $t_{d}$ are the time-scale for the rising and decaying phases of an outburst, respectively. 

For the fits of the two outbursts, $F_{c}$ is the weighted average of the signal over the entire observation period shown in Figure \ref{fig_fred}. Given the limited statistics for the first outburst O1, a two-stage fitting process was employed. In the first step, the peak time $t_{0}$ and the amplitude $F_{0}$ were determined using a gaussian fit, yielding $F_{0}=40.84$ and $t_{0}=59934.05$. In the second step, the rise and decay timescales were fitted as free parameters, resulting in $t_{r}=0.6\pm0.3$ days and $t_{d}=0.9\pm0.4$ days. For the second outburst O2 only $F_{c}$ is fixed, then $t_{r}=2.7\pm1.1$ days and $t_{d}=0.6\pm0.3$ days are estimated from the fitting.

Both of the two outbursts exhibit rising and decaying time-scale on the order of 1 day. For O1, the rise time-scale and decay time-scale are consistent within the uncertainties, whereas for O2 the decay time-scale is shorter than the rise time-scale. The detailed fitting results are provided in Table.\ref{tab:fitparameters}. Benefiting from its high-duty cycle and sensitivity, WCDA is capable of observing the complete evolution of both outbursts over the span of a month.

The variability time-scale provides an opportunity to estimate the size of the gamma-ray emitting region. Based on the observed time-scale, the (co-moving) radius of the production region at redshift z can be estimated using the formula $R'\leq ct_{var}\delta_{\rm D} (1+z)^{-1}$, where $\delta_{\rm D}$ represents the Doppler factor that describes relativistic beaming. By taking the rising time of O1 and O2 as $t_{r}=0.6\pm0.3$ days and $t_{r}=2.7\pm1.1$ days, respectively, the corresponding radius of the emission source should be $R'/\delta\leq 5.1\times 10^{14}$ cm and $\leq 6.8\times10^{15}$ cm, respectively. Naturally, both values are two to three orders of magnitude smaller than the size proposed in the 2009 flaring event paper of NGC 1275, which exhibited a time-scale on the order of months \citep{2014A&A...564A...5A}.

\begin{table}[h]
\centering
\caption{Fitting results for two outburst events}
\resizebox{\textwidth}{!}{%
\begin{tabular}{cccccc}
\hline
\hline
Epoch & $\chi^{2}/dof$ & $t_{0}$ & $F_{0}$ & $t_{r}$ & $t_{d}$\\
\hline
O1 & 1.0 & 59934.05 (fixed) & 40.84 (fixed) & $0.6\pm0.3$ & $0.9\pm0.4$\\
O2 & 0.9 & $59956.14\pm0.76$ & $53.29\pm16.65$ & $2.7\pm1.1$ & $0.6\pm0.3$\\
\hline
\hline
\end{tabular}%
}
\label{tab:fitparameters}
\end{table}

\subsection{Correlation analysis}

\begin{figure*}
\centering
\includegraphics[width=0.9\textwidth]
{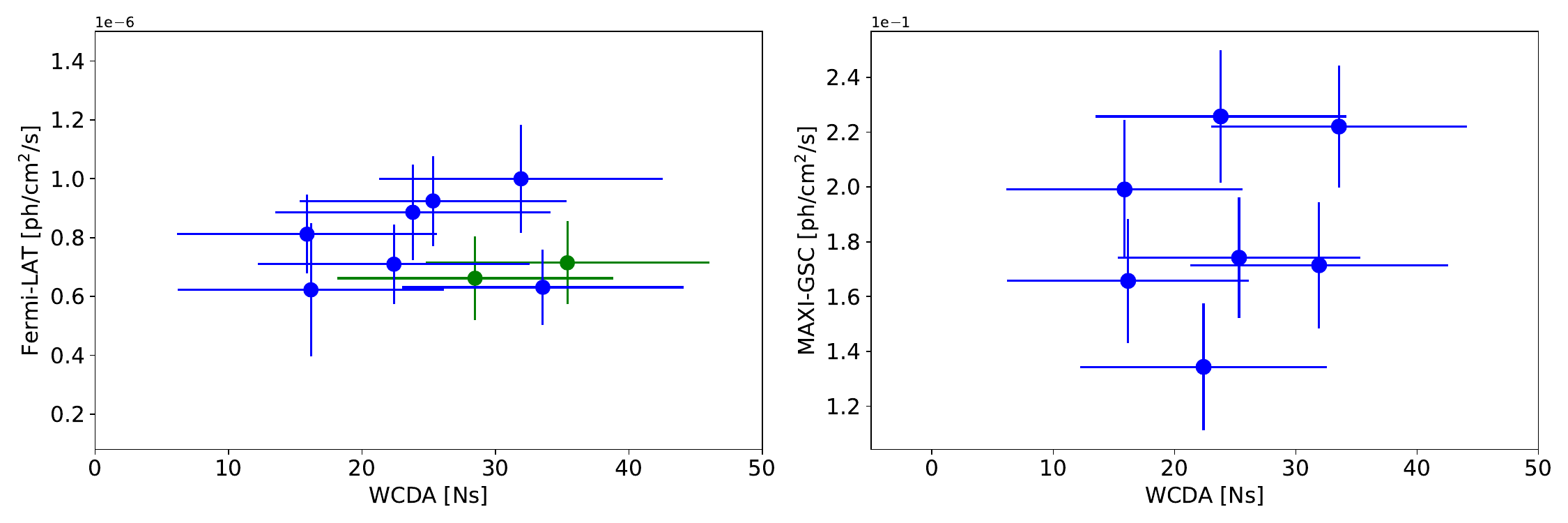}
\caption{Correlation analysis across different energy bands during outburst phases. 
The green points indicate the O1 period, and the blue points represent the O2 period. The left panel shows the correlation between LHAASO-WCDA and Fermi-LAT data, while the right panel displays the correlation between LHAASO-WCDA and MAXI-GSC data.} 
\label{fig_correlation}
\end{figure*}

Figure \ref{fig_correlation} illustrates the flux correlations between different energy bands for both outbursts O1 and O2. The left panel shows the correlation between the TeV and GeV bands, while the right panel focuses on the correlation between the TeV and keV bands. In this figure, the green and blue colors represent the outbursts O1 and O2, respectively.  
We utilized Pearson’s product-moment correlation coefficient, $r_{p}$, a commonly employed metric to assess the linear relationship between two variables, to determine the strength of correlations between different energy bands. Additionally, we also used Spearman’s rank correlation coefficient, $r_{sp}$, which is effective in evaluating nonlinear relationships.
In the left panel of Figure \ref{fig_correlation}, a correlation between the WCDA and Fermi-LAT for O1 and O2 reveals a correlation between the GeV and TeV bands. The correlation coefficients are noted as $r_p$ = 0.05 (p-value = 0.89) and $r_{sp}$ = 0.08 (p-value = 0.83).
In the right panel of Figure \ref{fig_correlation}, the correlation between soft X-ray and TeV gamma-ray bands is relatively weak, with correlation coefficients of $r_p$ = 0.25 (p-value = 0.58) and $r_{sp}$ = 0.29 (p-value = 0.53). This modest positive correlation suggests that the relationship between these two energy bands is not statistically significant. 
The overall results suggest that the correlation between the TeV and soft X-ray bands is slightly stronger than that between the TeV and GeV bands, but none of the correlations are statistically significant. Thus, the possibility that these emissions have different origins cannot be excluded.

\subsection{Significance map}

\begin{figure*}
\centering
\includegraphics[width=0.9\textwidth]{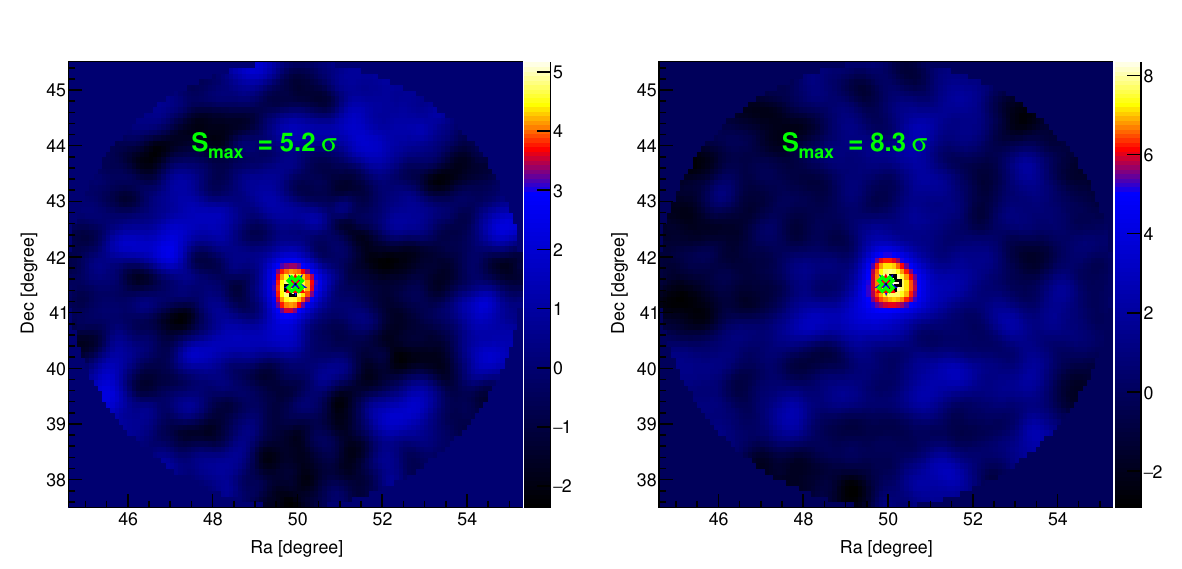}
\caption{Significance map of NGC 1275, highlighting the comparison between two specific observational periods, O1 and O2, displayed on the left and right panels. The black hollow cross marks the best-fit location from LHAASO-WCDA data, the blue hollow star indicates the position provided by Fermi-LAT, the red hollow triangle represents the position provided by MAXI-GSC, and the green hollow cross denotes the position in the optical band.}
\label{fig_sig}
\end{figure*}

The significance of two identified outbursts from NGC 1275 has been calculated using the Li-Ma prescription \citep{Li:1983fv}, and the results are illustrated in Figure \ref{fig_sig}. At energies above 1~\rm TeV, the emission from NGC 1275 is detected with a significance of $5.2\sigma$ during O1 period, the best-fit position derived through data is R.A. = $49.89^{\circ}\pm0.15^{\circ}$ and Dec. = $41.44^{\circ}\pm0.09^{\circ}$.

Similarly, during the O2 period, at energies higher than 1~\rm TeV, a signal significance of $8.3\sigma$ is achieved with the best-fit position of 
R.A. = $50.05^{\circ}\pm0.15^{\circ}$ and Dec. = $41.64^{\circ}\pm0.09^{\circ}$. For both outbursts, the observed centroid positions are consistent with the location of NGC 1275 with a distance of $0.1^{\circ}$. Additionally, it is also worth noting that for both O1 and O2, no photons with energy exceeding 3~TeV were detected. 

Expanding on this analysis, we conducted a posterior significance evaluation to further confirm the reliability of O1 and O2. We generated 10,000 light curves containing only background noise and identified the maximum significance value across multiple time windows for each light curve. By integrating and fitting the distribution of maximum significance, we determined the number of independent trials, thereby correcting for biases associated with multiple testing\citep{2015PhDT........91Z}. Utilizing the observed significance of O1 and O2, we calculated the prior probability and adjusted it with the count of independent trials to establish the posterior probability. The findings revealed a posterior significance of $4.4\sigma$ for O1 and $7.8\sigma$ for O2, indicating that both flares are unlikely to be attributed to background fluctuations.

\begin{figure*}
\centering
\includegraphics[width=0.9\textwidth]{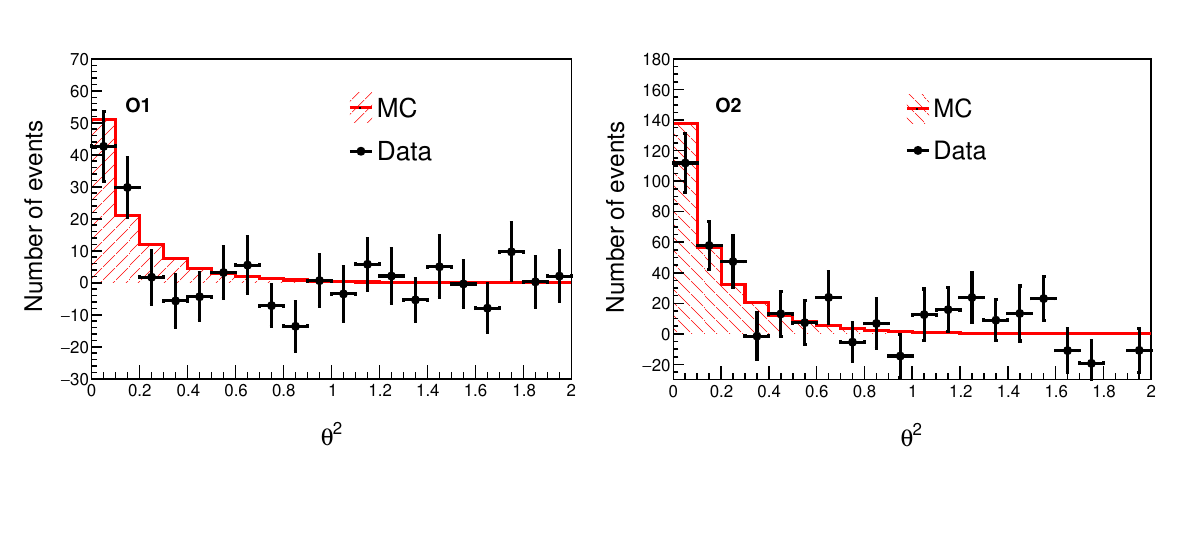}
\caption{Distribution of excess events as a function of the squared space angle relative to the NGC 1275 for experimental data and Monte Carlo (MC) simulations during the two outburst states.}
\label{fig_excess}
\end{figure*}

Figure \ref{fig_excess} compares the excess event distribution above 1~\rm TeV as a function of the squared distance from NGC 1275 ($\theta^2$) with PSF obtained from the Crab Nebula. 
The shape of the detected signal perfectly matches that of a point-like source, as expected for an AGN. 
This suggests that no detectable diffuse gamma-ray structures are present within the cluster during these two active periods.

\subsection{Spectral analysis}

\begin{figure*}
\centering
\includegraphics[width=0.9\textwidth]{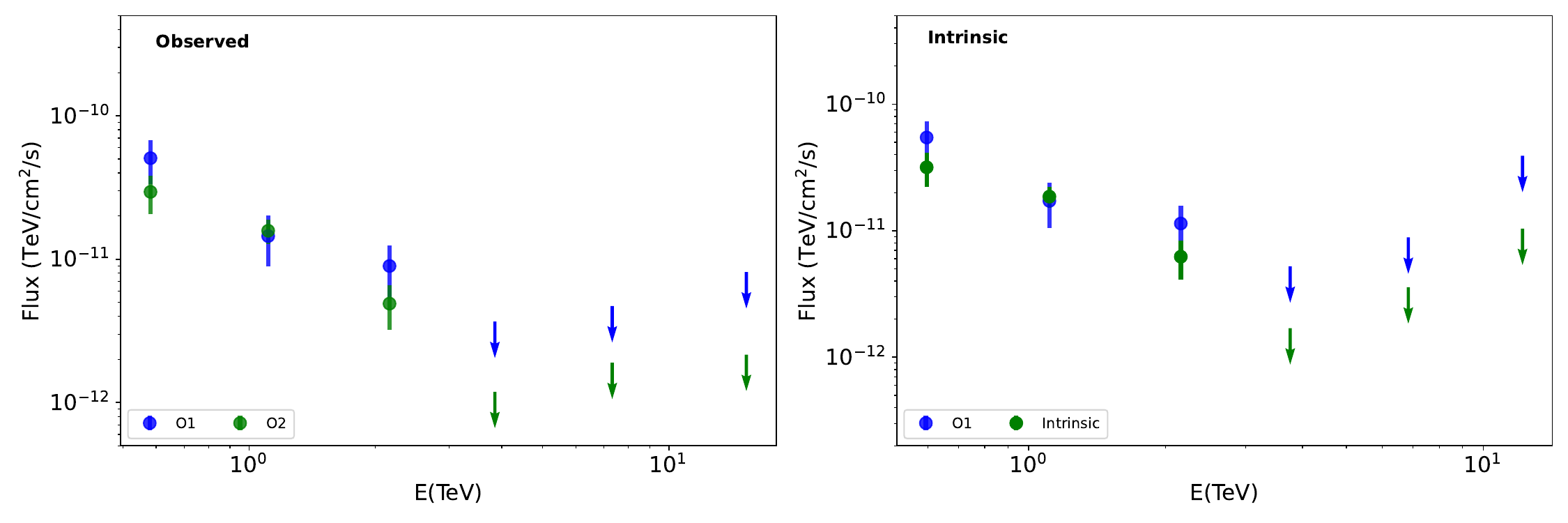}
\caption{The SED of NGC1275: left panel is for observed SEDs for the two outbursts, right panel is for intrinsic SEDs.}
\label{fig_SED}
\end{figure*}

The SED is estimated based on a forward-folding method \citep{aharonianPerformanceLHAASOWCDAObservation2021a}. The expected number of events in each energy bin is calculated by simulating the detector’s response to each signal event and assuming a power-law spectrum defined as $dN/dE=\phi_0(E/E_0)^{-\alpha}$, the referenced $E_{0}$ is chosen to be 3~\rm TeV. The best-fit values of ($\phi_0$, $\alpha$) are obtained by minimizing $\chi^2$ function.

In this study, the observed differential energy spectrum spanning from 500 GeV to 3 TeV is well described by a power-law model. The best-fit parameters for the first outburst, O1, are $\phi_{0}=(5.15\pm 3.04)\times 10^{-13} ~\rm TeV^{-1}cm^{-2}s^{-1}$, $\alpha=-3.37\pm 0.52$, with a $\chi^{2}/{\rm dof}$ = 0.8. For the second outburst, O2, the parameters are $\phi_{0}=(4.01\pm 1.32)\times 10^{-13} ~\rm TeV^{-1}cm^{-2}s^{-1}$ and $\alpha=-3.35\pm 0.29$, with a $\chi^{2}/{\rm dof}$ = 0.78. Here, the total systematic uncertainty of the flux is approximately 8\%-24\% \citep{2023SciA....9J2778C}. 
The measured integral flux between 0.5 to 3 TeV is $4.38\times10^{-11} ~\rm cm^{-2}s^{-1}$ and $3.31\times10^{-11} ~\rm cm^{-2}s^{-1}$, corresponding to about 64\% and 48\% of the Crab Nebula flux. 
Remarkably, the flux levels during these two outburst phases are exceptionally high, more than 200 times, compared to the long-term average spectral energy distribution measured by 253 hours of MAGIC observation spanning from 2009 to 2014 \citep{2016A&A...589A..33A}.

Since the source is of extragalactic origin, 
the estimation of the intrinsic energy spectrum takes into consideration the absorption effects of the extragalactic Background Light (EBL).
The input spectral model assumes a form of $\phi(E)=\phi_0(E/E_0)^{-\alpha}\times {\rm exp}(-\tau(E))$. Here, the EBL model utilized for fitting is sourced from \citet{2021MNRAS.507.5144S}. This effect can be approximated by a decrease (hardening) in the power-law index of about 0.12 and a 30\% increase in the differential flux normalization at 3~\rm TeV. Both the observed and the intrinsic differential energy spectra are depicted in Figure \ref{fig_SED}, with the detailed fit parameters presented in Table \ref{tab_sedfit}.

\begin{table}
\centering
\begin{tabular}{ccc}
\hline
\hline
Epoch & $\phi_{0}$ & $\alpha$\\
  & $ (10^{-13} TeV^{-1}cm^{-2}s^{-1}$) & \\
\hline
O1 (observed) & $5.15\pm3.04$ &$-3.37\pm0.52$\\
O2 (observed) & $4.01\pm1.32$ &$-3.35\pm0.29$ \\
\hline
O1 (intrinsic) & $6.85\pm4.07$ &$-3.25\pm0.53$\\
O2 (intrinsic) & $5.25\pm1.75$ &$-3.25\pm0.29$ \\
\hline 
\hline
\end{tabular}
\caption{The fitting parameters of the LHAASO-WCDA SEDs.}
\label{tab_sedfit}
\end{table}

\begin{figure*}
\centering
\includegraphics[width=0.9\linewidth]{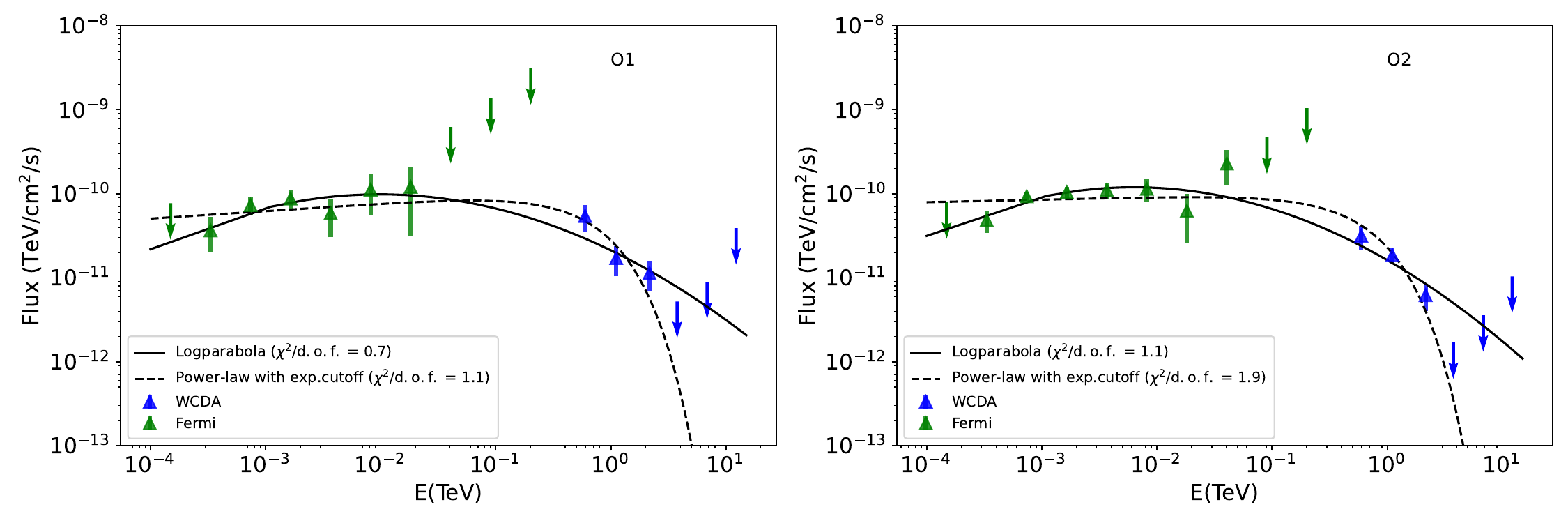}
\caption{The SED in gamma-ray energy band of NGC 1275. The left panel displays the fitting results for O1, the right panel is the fitting results for O2.}
\label{fig_sedfit}
\end{figure*}

In Figure \ref{fig_sedfit}, the EBL-corrected SED measured by WCDA ranging from 0.5 to 3~\rm TeV is compared with the simultaneously measured energy spectrum from Fermi-LAT spanning 0.2 - 20~\rm GeV. The comparison suggests a notable steepening of the spectrum. In order to investigate the transition between Fermi-LAT data and WCDA data, we fitted two functions to these gamma-ray SEDs, 
taking into account both the statistical and systematic errors of each data point. Table \ref{tab:fit_parameter} gives the function norms used and the fit results. 

Analyzing the Figure \ref{fig_sedfit}, we observed the log-parabola fit ($dF/dE=F_{0}{\times}E^{-\alpha-\beta \log(E)}$) offers a more accurate fit to the Fermi-LAT data, achieving a lower chi-squared value particularly in the low-energy region, compared to the exponential cutoff power-law model ($dF/dE=F_0\times E^{-\alpha}e^{-E/E_{c}}$). However, the log-parabola model tends to overestimate the flux above 3 ~\rm TeV (see Fig.~\ref{fig_sedfit}), becoming incompatible with the WCDA upper limits. Consequently, we prefer to adopt the power-law with an exponential cutoff function to describe these two outbursts, as it provides a more consistent representation of the observed spectra over the entire energy range.

It allows us to determine the position of the cutoff energy of the SED from two campaigns both in full sky survey mode, which shows no large difference and remains consistent within the uncertainties of $0.70\pm0.40$ TeV and $0.66\pm0.16$ TeV. Interestingly, this cutoff energy position is in good agreement with the value of $0.49\pm0.35$ \citep{2018A&A...617A..91M} provided by the joint analysis of MAGIC and Fermi-LAT for the Jan. 2017 NGC 1275 flaring events.

\begin{table*}
\centering
\caption{Parameters of the joint spectral fit to the gamma-ray SEDs obtained with LHAASO-WCDA and Fermi-LAT data for two fitting functions (with $1\sigma$ uncertainties). }
\begin{tabular}{lcccc}
\hline
\hline
\multicolumn{5}{c}{Log parabola : $dF/dE = f_{0} \times E^{-\alpha-\beta \times \log(E)}$} \\
\cline{1-5}
Epoch & $F_{0}$ & $\alpha$ & $\beta$ & $\chi^2/dof $ \\
& $(\times10^{-11}TeV^{-1}cm^{-2}s^{-1})$ &&&\\
\hline
O1 & $2.12\pm0.46$ & $2.66\pm 0.15$ & $0.07\pm 0.02$ & 0.7\\
O2 & $1.63\pm0.25$ & $2.79\pm 0.09$ & $0.08\pm 0.01$ & 1.1\\
\hline
\hline
\end{tabular}

\begin{tabular}{lcccc}
\hline
\hline
\multicolumn{5}{c}{Power-law with exponential cutoff : $dF/dE = f_{0} \times E^{-\alpha}\times e^{-E/E_{c}}$} \\
\cline{1-5}
Epoch & $F_{0}$ & $\alpha$ & $E_{c}$ & $\chi^2/dof $ \\
& $(\times10^{-10}TeV^{-1}cm^{-2}s^{-1})$&&(TeV)&\\
\hline
O1 & $1.17\pm0.95$ & $1.91\pm0.12$ & $0.70\pm0.40$ & 1.1 \\
O2 & $1.06\pm0.39$ & $1.97\pm0.05$ & $0.65\pm0.16$ & 1.9 \\
\hline
\hline 
\end{tabular}

\label{tab:fit_parameter}
\end{table*}

\subsection{Modelling test to the gamma-ray emission of NGC 1275}

\begin{figure*}
\centering
\includegraphics[width=0.9\linewidth]{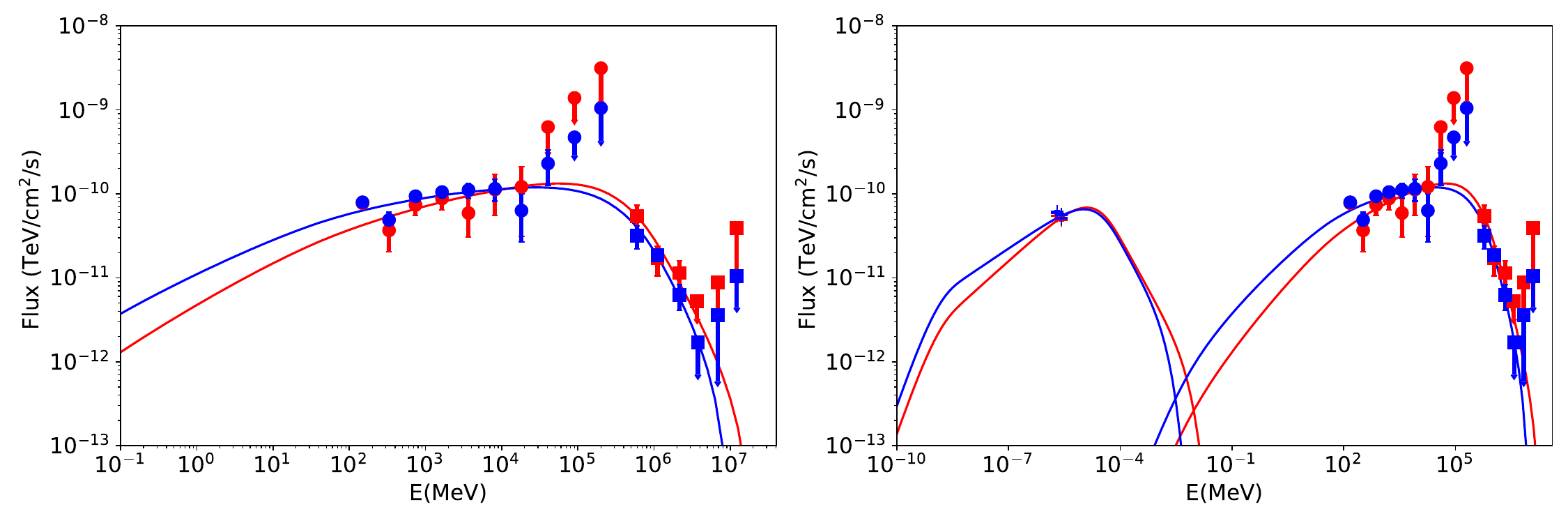}
\caption{The left panel displays the gamma-ray spectral energy distribution of NGC 1275, with the red and blue lines representing the SSC fits for the O1 and O2 outbursts, respectively. The right panel combines measurements the optical band during outbursts, along with the corresponding SSC fits for O1 (red) and O2 (blue).} 
\label{fig:sedssc}
\end{figure*}

We employed the one-zone SSC model \citep{2008ApJ...686..181F}, 
to reproduce VHE emission for each of the outbursts from NGC 1275. The model considers a homogeneous spherical region or blob with radius $R_b^\prime$ in the jet co-moving frame with a Doppler factor $\delta_D$ 
and a randomly oriented magnetic field of mean intensity $B^\prime$. 
We then consider the electron energy distribution in the emission region to be a broken power-law as
\begin{equation}
\begin{aligned}
N_e^\prime(\gamma^\prime) &= K_e^\prime\left[\left(\frac{\gamma^\prime}{\gamma_b^\prime}\right)^{-p_1}H(\gamma_b^\prime-\gamma^\prime)\right.\\
&\quad +\left.\left(\frac{\gamma^\prime}{\gamma_b^\prime}\right)^{-p_2}H(\gamma^\prime-\gamma_b^\prime)\right]H(\gamma^\prime;\gamma_1^\prime,\gamma_2^\prime),
\end{aligned}
\end{equation}
where $H(x)=0$ for $x<0$ and $H(x)=1$ for $x \ge 0$, as well as $H(x;x_1,x_2)=1$ for $x_1 \leq x \leq x_2$ and $H(x;x_1,x_2)=0$ everywhere else, $\gamma^\prime$ is the electron Lorentz factor, $p_1$ and $p_2$ are the power-law indices, $\gamma_b^\prime$is the break electron Lorentz factor, $K_e^\prime$ is the normalization constant, $\gamma_1^\prime$ and $\gamma_2^\prime$ are the minimum and maximum electron Lorentz factor. 
In Figure \ref{fig_sedfit}, we only fitted the gamma-ray SED with a power-law with an exponential cutoff function but did not consider the physical processes producing these spectra. The fitting result seems to be quite accurate near the center of the spectrum but loses accuracy at low and high frequencies. In order to improve the accuracy of the fit for SED, here we use a broken-power-law model for energetic electrons to fit the observational multiwavelength data. We adopt this model because fast cooling of the higher energetic electrons generally produces broken power-law distributions of the emitting particles, during rapid events like flares.

Figure \ref{fig:sedssc} demonstrates the SED modeling 
of NGC 1275, which is consistent with the observational data in the gamma-ray band. The X-ray emission is likely contaminated by the cluster emission, hence the MAXI data was omitted from the SED. 
The break Lorentz factor of the electron energy distribution is $\gamma'_b=1.5\times 10^5$. The electron spectral indices are $p_1=2.2$ and $p_2=4.6$. The magnetic field strength is $B'=2.5$ mG. The Doppler factor is 20. In this case, the synchrotron peak of the broad-band SED is in the optical band. 


NGC 1275 has been observed in multi-wavelength campaigns. However, we note that the source is extended and the morphology is complex \citep{2021ApJ...906...30I}. The jet of NGC 1275 interacts with the surrounding medium. The TeV emission is expected from either the jet itself or the interaction region. 
In this paper, we perform the test using a one-zone SSC model for the gamma-ray emission of NGC 1275 detected by LHAASO-WCDA. The optical depth of the gamma-ray emission is significantly smaller than 1 \citep{2008ApJ...686..181F}. It implies that the flux decrease above 1 TeV is not caused by intrinsic absorption. It is shown that the lepton model can successfully explain the VHE emission of NGC 1275 detected by LHAASO-WCDA. 
The modeling of multi-component radiation can be comprehensively studied in the future.

\section{Summary and conclusion} 
\label{sec:discuss}

In this paper, we conducted a comprehensive investigation into the temporal behavior, morphology, and spectral characteristics of NGC 1275 based on detections by LHAASO-WCDA during Nov. 2022 and Jan. 2023. The high-duty cycle of the detector enabled us to observe the complete evolution of two outbursts, encompassing distinct rising and decay phases. 
The accumulated significance reaches 5.2~$\sigma$ and 8.3~$\sigma$ for the two outbursts. From our analysis, the following conclusions can be drawn:

\begin{itemize}

\item No apparent change in the gamma-ray spectral index has been observed during two flaring periods, remaining around 3.4 within the uncertainties. This may imply that the same process is responsible for these two outbursts. 

\item The flux in the energy range above 0.5 TeV during two outbursts reached approximately 64\% and 48\% of the Crab Nebula flux, respectively. These flux levels are exceptionally high, indicating an increase by factors of approximately 300 and 200 times compared to the flux measured during 253 hours of long-term observation by MAGIC on the same source\citep{2016A&A...589A..33A}.

\item Our observation has revealed that NGC 1275 exhibits variability in energies higher than 0.5~TeV on time-scale of days. For the first outburst (O1), the rising and decay time-scale ($\tau_r,\tau_d$) are ($0.6\pm0.3$ days, $0.9\pm0.4$ days), while for the second outburst (O2), they are ($2.7\pm1.1$ days, $0.6\pm0.3$ days).

\item The shape of the excess signal closely resembles that of a point-like source, suggesting the lack of any noticeable diffuse gamma-ray structure within the cluster.

\item Based on simultaneous Fermi-LAT data, GeV-to-TeV gamma-rays can be well described with a power-law with an exponential cut function. The position of the cutoff energy is approximately $0.70\pm0.40~\rm TeV$ and $0.65\pm0.16~\rm TeV$, which is consistent with the value, $0.49\pm0.35~\rm TeV$, observed during the NGC 1275 flare in 2017 by MAGIC.

\item We adopt the one-zone SSC as the dominant mechanism to explain the VHE emission of NGC 1275 observed by LHAASO-WCDA. The emission region is constrained to be less than $1.6\times 10^{17}$ cm from the VHE variability time-scale. The acceleration is efficient allowing electrons to reach $\gamma\sim 10^6$ for the radiation, and the magnetic field strength in the VHE emission region is as weak as 2.5 mG. In such a case, it indicates that the VHE production region is electron-energy dominated (not magnetic-energy dominated).

\end{itemize}


The detection of the outburst behavior in NGC 1275 by LHAASO-WCDA provided detailed insights into its spectral, morphological, and temporal characteristics. During the outburst, the gamma-ray spectral index remained stable, while the flux in the high-energy range significantly increased. These findings indicate that the high-energy radiation processes in NGC 1275 are likely driven by consistent physical mechanisms, especially during periods of intense activity, offering valuable insights into the study of relativistic jets in active galactic nuclei.

\section{Acknowledgements}
We would like to thank all staff members who work at the LHAASO site above 4400\,m a.s.l. year round to maintain the detector and keep the water recycling system, electricity power supply and other components of the experiment operating smoothly. We are grateful to Chengdu Management Committee of Tianfu New Area for the constant financial support for research with LHAASO data. This research work is also supported by the following grants: the National Natural Science Foundation of China (NSFC grants No. 12393853, No. 12393854, No. 12393851, No. 12393852, No. 12175121, No. 12173066, No. 12173039, No. 12393813), the National Key R\&D Program of China (2023YFE0101200), the Department of Science and Technology of Sichuan Province, China (Grant No. 24NSFSC0449) and in Thailand by the National Science and Technology Development Agency (NSTDA) and National Research Council of Thailand (NRCT): High-Potential Research Team Grant Program (N42A650868). J.R. Mao is also supported by the Yunnan Revitalization Talent Support Program (YunLing Scholar Project).

The optical data of this work is from ZTF. 
ZTF is supported by the National Science Foundation under grants Nos. AST-1440341 and AST-2034437 and a collaboration including current partners Caltech, IPAC, the Weizmann Institute for
Science, The Oskar Klein Center at Stockholm University, the
University of Maryland, Deutsches Elektronen-Synchrotron
and Humboldt University, the TANGO Consortium of Taiwan,
the University of Wisconsin at Milwaukee, Trinity College
Dublin, Lawrence Livermore National Laboratories, IN2P3,
University of Warwick, Ruhr University Bochum, Northwestern
University and former partners the University of
Washington, Los Alamos National Laboratories, and Lawrence
Berkeley National Laboratories. Operations are conducted by
COO, IPAC, and UW. This research has made use of the
NASA/IPAC Extragalactic Database (NED), which is operated
by Jet Propulsion Laboratory, California Institute of Technology,
under contract with the National Aeronautics and Space
Administration.
Facilities: Fermi, Swift, WISE.
Software: astropy\citep{2013A&A...558A..33A, 2018AJ....156..123A, 2022ApJ...935..167A}, naima\citep{Zabalza:2015bsa}.

\section{Author Contribution}
M. Zha and J.R. Mao led the drafting of text and coordinated the whole data analysis, R. Wang and C.F. Feng performed the data analysis, C.Y. Yang and J.R. Mao contributed to the interpretation of the data and modeling of SEDs, C.D. Gao provided the cross-check. All other authors participated in data analysis, including detector calibration, data processing, event reconstruction, data quality check, and various simulations, and provided comments on the manuscript.

\section{Data Availability}
The Fermi-LAT data used in this work is publicly available, which is provided online by the NASA-GSFC Support Center\footnote{\url{ https://fermi.gsfc.nasa.gov/ssc/data/access/lat/}}.
The MAXI-GSC data used in this work were calculated based on the on-demand service provided by MAXI.\footnote{\url{http://maxi.riken.jp/mxondem/}}. 
The ZTF data in g and r bands are collected from the 20th ZTF public data release \footnote{\url{https://irsa.ipac.caltech.edu/cgi-bin/Gator/nph-scan}}. The LHAASO data underlying this article will be shared on reasonable request to the corresponding author.





\bibliographystyle{mnras}
\bibliography{NGC1275_template} 








\twocolumn

\vspace*{0.5cm}
\noindent $^{1}$ Key Laboratory of Particle Astrophysics \& Experimental Physics Division \& Computing Center, Institute of High Energy Physics, Chinese Academy of Sciences, 100049 Beijing, China\\
$^{2}$ University of Chinese Academy of Sciences, 100049 Beijing, China\\
$^{3}$ TIANFU Cosmic Ray Research Center, Chengdu, Sichuan,  China\\
$^{4}$ Dublin Institute for Advanced Studies, 31 Fitzwilliam Place, 2 Dublin, Ireland \\
$^{5}$ Max-Planck-Institut for Nuclear Physics, P.O. Box 103980, 69029  Heidelberg, Germany\\
$^{6}$ School of Physical Science and Technology \&  School of Information Science and Technology, Southwest Jiaotong University, 610031 Chengdu, Sichuan, China\\
$^{7}$ School of Astronomy and Space Science, Nanjing University, 210023 Nanjing, Jiangsu, China\\
$^{8}$ Center for Astrophysics, Guangzhou University, 510006 Guangzhou, Guangdong, China\\
$^{9}$ Tsung-Dao Lee Institute \& School of Physics and Astronomy, Shanghai Jiao Tong University, 200240 Shanghai, China\\
$^{10}$ Institute for Nuclear Research of Russian Academy of Sciences, 117312 Moscow, Russia\\
$^{11}$ Hebei Normal University, 050024 Shijiazhuang, Hebei, China\\
$^{12}$ University of Science and Technology of China, 230026 Hefei, Anhui, China\\
$^{13}$ State Key Laboratory of Particle Detection and Electronics, China\\
$^{14}$ Key Laboratory of Dark Matter and Space Astronomy \& Key Laboratory of Radio Astronomy, Purple Mountain Observatory, Chinese Academy of Sciences, 210023 Nanjing, Jiangsu, China\\
$^{15}$ Research Center for Astronomical Computing, Zhejiang Laboratory, 311121 Hangzhou, Zhejiang, China\\
$^{16}$ Key Laboratory for Research in Galaxies and Cosmology, Shanghai Astronomical Observatory, Chinese Academy of Sciences, 200030 Shanghai, China\\
$^{17}$ School of Physics and Astronomy, Yunnan University, 650091 Kunming, Yunnan, China\\
$^{18}$ Key Laboratory of Cosmic Rays (Tibet University), Ministry of Education, 850000 Lhasa, Tibet, China\\
$^{19}$ Department of Physics, The Chinese University of Hong Kong, Shatin, New Territories, Hong Kong, China\\
$^{20}$ Key Laboratory of Radio Astronomy and Technology, National Astronomical Observatories, Chinese Academy of Sciences, 100101 Beijing, China\\
$^{21}$ School of Physics and Astronomy (Zhuhai) \& School of Physics (Guangzhou) \& Sino-French Institute of Nuclear Engineering and Technology (Zhuhai), Sun Yat-sen University, 519000 Zhuhai \& 510275 Guangzhou, Guangdong, China\\
$^{22}$ Institute of Frontier and Interdisciplinary Science, Shandong University, 266237 Qingdao, Shandong, China\\
$^{23}$ APC, Universit\'e Paris Cit\'e, CNRS/IN2P3, CEA/IRFU, Observatoire de Paris, 119 75205 Paris, France\\
$^{24}$ Department of Engineering Physics \& Department of Astronomy, Tsinghua University, 100084 Beijing, China\\
$^{25}$ School of Physics and Microelectronics, Zhengzhou University, 450001 Zhengzhou, Henan, China\\
$^{26}$ Yunnan Observatories, Chinese Academy of Sciences, 650216 Kunming, Yunnan, China\\
$^{27}$ China Center of Advanced Science and Technology, Beijing 100190, China\\
$^{28}$ College of Physics, Sichuan University, 610065 Chengdu, Sichuan, China\\
$^{29}$ School of Physics, Peking University, 100871 Beijing, China\\
$^{30}$ Guangxi Key Laboratory for Relativistic Astrophysics, School of Physical Science and Technology, Guangxi University, 530004 Nanning, Guangxi, China\\
$^{31}$ Department of Physics, Faculty of Science, Mahidol University, Bangkok 10400, Thailand\\
$^{32}$ Moscow Institute of Physics and Technology, 141700 Moscow, Russia\\
$^{33}$ Center for Relativistic Astrophysics and High Energy Physics, School of Physics and Materials Science \& Institute of Space Science and Technology, Nanchang University, 330031 Nanchang, Jiangxi, China\\
$^{34}$ National Space Science Center, Chinese Academy of Sciences, 100190 Beijing, China\\
$^{35}$ School of Physics, Huazhong University of Science and Technology, Wuhan 430074, Hubei, China\\
\bsp	
\label{lastpage}
\end{document}